\documentclass[]{aa}
\usepackage{natbib}
\usepackage{txfonts}
\usepackage{graphicx}

\bibpunct{(}{)}{;}{a}{}{,} 

\def\kms{$\mathrm{km\, s^{-1}}$ }

\def\Msun{\hbox{$\hbox{M}_\odot$}}

\begin{document}

\title{The isotopic $\rm ^{6}Li/^{7}Li$ ratio
in Cen X-4 and the origin of Li in X-ray binaries
\thanks {based on observations collected at ESO - Paranal 
in programmes 065-0447 and 073-0214}}

\author{J.Casares\inst{1}
\and
P.Bonifacio\inst{2,3,4}\and J.I. Gonz\'alez Hern\'andez\inst{1,2,3} 
 \and P.Molaro\inst{3} \and M. Zoccali\inst{4}
}
\institute{Instituto de Astrof\'\i{}sica de Canarias, E-38200 La
Laguna, Tenerife, Spain\\
\email{jcv@ll.iac.es}
\and
CIFIST Marie Curie Excellence Team
\and
Observatoire de Paris, GEPI, F-92195 Meudon Cedex, France\\
\email{Piercarlo.Bonifacio@obspm.fr,Jonay.Gonzalez-Hernandez@obspm.fr}
\and
Istituto Nazionale di Astrofisica - Osservatorio Astronomico di
Trieste,  Via Tiepolo 11, I-34143  Trieste, Italy\\
\email{molaro@ts.astro.it}
\and
Pontificia Universitad Catolica de Chile\\
\email{mzoccali@puc.cl} 
}

\authorrunning{Casares et al.}
\titlerunning{UVES spectroscopy of Cen X-4}

\offprints{J. Casares}
\date{Received xxx; Accepted xxx}

\abstract
{Cool stars, companions to compact objects, are known to show
Li abundances which are high compared to field stars of the same
spectral type, which are heavily Li depleted. This may be due
either to Li production  or Li preservation
in theses systems.}
{
To measure  the lithium isotopic ratio in the companion star of the 
neutron star X-ray binary Cen X-4.} 
{We use
 UVES spectra obtained in years 
2000 and 2004 around the orbital quadratures.
The spectra are analysed with spectrum synthesis
techniques and the errors estimated with
Monte Carlo simulations.}
{
We measure A(Li)=$2.87\pm0.20$ and $^6$Li/$^7$Li = 
$0.12^{+0.08}_{-0.05}$ at 68\% confidence level. We also
present updated system parameters with a refined determination of the 
orbital period and component masses i.e. $1.14\pm0.45$ \Msun~and 
$0.23\pm0.10$ \Msun~for the neutron star and companion, respectively.
}
{
In our view the low level of $^6$Li favours Li preservation scenarios,
although Li production mechanisms cannot be ruled out. 
In the case of preservation, no Li is freshly created in the binary,
but the tidally-locked companion has preserved its original Li by 
some mechanism, possibly inhibited destruction due to its fast
rotation.  
}
\maketitle
\markboth{Cen X-4}{}

\section{Introduction}

Spectroscopic analysis of the companion stars to X-ray binaries provides
unique information on the nature of the accreting compact
objects (e.g. Casares 2001) and their formation history (Israelian et al. 
1999, Gonz\'alez Hern\'andez et al. 2004). The subclass of Soft
X-ray Transients (SXTs) provide excellent testbeds due to
long periods of quiescence during which  the companion star is not
overwhelmed by the (otherwise) intense X-ray reprocessed disc emission 
(van Paradijs \& McClintock 1995).
The study of SXTs has led to the discovery of a large population of
black holes in the Galaxy, recognized by their large mass
functions (e.g. Charles \& Coe 2006, Casares 2005) exceeding 3 \Msun.
A sizable fraction of SXTs have shown Type-I X-ray bursts and,
therefore, they contain accreting neutron stars.  

Cen X-4 is the nearest member of the class of neutron star SXTs with two 
X-ray outbursts observed in 1969 (Conner, Evans \& Belian 1969) and 1979 
(Kaluzienski, Holt \& Swank 1980). V822 Cen, the optical counterpart of 
Cen X-4, was discovered in the course of the 1979 outburst (Canizares, 
McClintock \& Grindlay 1980). Since 1980 it has stayed in a quiescent level 
at V$\simeq$18.2 and it 
has been the target of intensive investigation which led to the discovery 
of the 15.1 hr orbital period, through photometric 
modulations (Chevalier et al. 1989), and the determination of the
companion's radial velocity curve (Cowley et al. 1988; McClintock \& 
Remillard 1990; Torres et al. 2002). The 
different orbital solutions show a consistent velocity semi-amplitude
($K_2 = 146-150$ km s$^{-1}$) but a large scatter in systemic velocities which 
tend to converge over the years from an inital 137-234 km s$^{-1}$ (Cowley et 
al. 1988, McClintock \& Remillard 1990) to 184-196 km s$^{-1}$ (Torres et al. 
2002, D'Avanzo et al. 2005). The large systemic velocity and high galactic 
latitude ($b=+24^{\rm o}$) is consistent with Cen X-4 being a member Galaxy 
Halo population (Cowley et al. 1988). However, by integrating the 
binary motion in the Galactic gravitational potential, Gonz\'alez 
Hern\'andez et al. (2005a) propose that the binary was formed in the Galactic 
plane and projected into its current position by a natal kick velocity during
the supernova explosion. The inclination angle has been constrained to 
$i=31^{\circ}-54^{\circ}$ (Shahbaz et al. 1993) and the binary mass ratio to 
$q=M_2/M_{\rm NS}=0.17 \pm 0.06$ (Torres et al. 2002) 
which leads to $M_{\rm NS} = 0.5-2.1$ \Msun~ and 
$M_2 = 0.04-0.58$ \Msun~ for the masses of the neutron star and its 
companion.

The companion star has been classified as a K3-K7 star which must be 
substantially evolved in order to fill its $\sim 0.6$ R$_{\odot}$  Roche 
lobe (see Shahbaz et al. 1993 and references therein).  
Detailed atmospheric parameters ($T_{\rm eff}=4500 \pm 100$, $\log g=3.9 
\pm 0.3$) and chemical abundances of the donor star were presented in 
Gonz\'alez Hern\'andez et al. (2005b). The low gravity provides additional 
support for an expanded atmosphere. Furthermore, a super solar metallicity 
[Fe/H]=0.23 $\pm$ 0.10 and moderately enhanced Ti and Ni is found, which 
supports contamination from supernova products.  
The companion also presents an anomalous $^{7}$Li overabundance for a K type 
star (Mart\'\i{}n et al. 1994; Gonz\'alez Hern\'andez et al. 2005b). This 
latter 
peculiarity, which is also common to the companions of several black hole 
binaries, has been explained by spallation reactions in the course of the X-ray
outbursts (Mart\'\i{}n et al. 1994), although other mechanisms for Li
production might be
possible (e.g. Yi \& Narayan 1997). Alternatively, the companion could have
preserved its primordial Li abundance if the binary were younger than 
$\simeq 10^7$ yr or depletion efficiently inhibited by the action of fast 
rotation (see Maccarone, Jonker \& Sils 2005). The origin of the large 
$^{7}$Li abundance could be addressed by measuring the isotopic ratio 
$^{6}$Li/$^{7}$Li and, therefore, we have embarked in a project to obtain high
resolution ($R\sim 40000$) spectroscopy of Cen X-4.

\section{Observations and reduction}

We obtained 20 spectra of Cen X-4 at the European Southern Observatory
(ESO), {\sl Observatorio Cerro Paranal}, using the 8.2m {\sl Very Large
Telescope} (VLT), equipped with the UVES echelle spectrograph, on the nights 
of 25 April and 9 June 2000, covering the spectral ranges 
$\lambda\lambda$4780--5755 (hereafter green) and $\lambda\lambda$5837--6808 
(hereafter red). A 1 
arcsec slit was used resulting in a resolving power R=43,000.   
The exposure time was fixed to 718 s and the observations were concentrated 
around the quadratures to minimize the effects of orbital smearing which, for 
the orbital parameters of Cen X-4, is in the range 0.2--10 km s$^{-1}$ i.e. 
smaller or comparable to our instrumental resolution of 7 km s$^{-1}$. 

Sixteen more spectra of Cen X-4 were obtained on four nights in 
April 2004 also with UVES. The instrumental set-up was identical to 
the 2000 observations except that the DIC1 dichroic was also employed, enabling 
us to obtain additional blue spectra covering $\lambda\lambda$3281--4562 
(hereafter blue). 
Unfortunately, the count level was very low for some spectra and only 9 could  
be included in the analysis. The exposure time was set to 665 s which, for the 
phase of our observations, results in an orbital smearing $\le 7$ km s$^{-1}$. 
A full observing log is presented in Table 1.

\begin{table}
\caption{Journal of Observations}
\label{tbl-1}
\begin{tabular}{ccccc}
\hline
Date & Wav. Range & Number & Exp. time & Orbital phase  \\
&   (\AA) & of spectra & (s.) & \\
\hline
\hline
25/4/2000  & 4780--5755  &  10 & 718 & 0.09-0.22\\
   ,,      & 5837--6808  &  ,, &  ,, &     ,,    \\
9/6/2000   & 4780--5755	 &  10 & 718 & 0.66-0.80\\
   ,,      & 5837--6808  &  ,, &  ,, &    ,,\\
3/4/2004   & 3281--4562  &   4 & 665 & 0.64-0.68 \\
   ,,      & 4780--5755  &  ,, &  ,, &   ,,   \\
   ,,      & 5837--6808  &  ,, &  ,, &   ,,   \\
12/4/2004  & 3281--4562  &   4 & 665 & 0.22-0.26 \\
   ,,      & 4780--5755  &  ,, &  ,, &   ,, \\
   ,,      & 5837--6808  &  ,, &  ,, &   ,, \\
16/4/2004  & 3281--4562  &   4 & 665 & 0.17-0.22 \\
   ,,      & 4780--5755  &  ,, &  ,, &   ,,  \\
   ,,      & 5837--6808  &  ,, &  ,, &   ,,  \\
21/4/2004  & 3281--4562  &   4 & 665 & 0.19-0.23 \\
   ,,      & 4780--5755  &  ,, &  ,, &   ,,\\
   ,,      & 5837--6808  &  ,, &  ,, &   ,, \\
\hline
\hline
\end{tabular}
\end{table}
  
We extracted the 1-dimensional spectra using optimal extraction within MIDAS 
ECHELLE context, and the pixel-to-wavelength calibration was obtained with the 
observation of a Th-Ar lamp. We compared our reduction with the ESO pipeline 
and found that they are comparable, except for the 2000 data where 
we prefer ours because it provides better statistics, in particular for the
green spectra. All the spectra were 
flux calibrated using nightly observations of ESO flux standards. 
We checked the accuracy of the wavelength
calibration using the $\lambda$6300.3 and $\lambda$5577.4 [OI] sky lines and 
found it to be precise within 0.4 km s$^{-1}$ for the two databases.

\section{Revised System Parameters}

Radial velocities of individual spectra were obtained through 
cross-correlating our red spectra ($\lambda\lambda$5837--6808) with a 
synthetic template computed for the 
stellar parameters and metallicity derived by Gonz\'alez Hern\'andez et al 
(2005b), i.e. $T_{\rm eff}=4500$, $\log g=3.9$, [Fe/H]=0.23, using Kurucz LTE 
models (Kurucz 1993). Prior to the cross-correlation, the synthetic 
spectrum was degraded to the resolution of our UVES data (7 km 
s$^{-1}$) and rotationally broadened by 44 km s$^{-1}$ to match the rotational 
velocity of the donor star in Cen X-4 (Torres et al. 2002). A spherical 
rotational profile with limb-darkening $\epsilon=0.65$ was employed 
(Gray 1992). The main disc emission lines (H$_{\alpha}$, HeI $\lambda$5875 
and $\lambda$6678) and interstellar/atmospheric absorptions (NaI D,
$\lambda$6300) were masked from the cross-correlation. A least-squares 
sine wave fit to our velocity points yields the following orbital solution: 

$$ \gamma= 189.6 \pm 0.2~ {\rm km s}^{-1}$$
$$P=0.6290522 \pm 0.0000004~ {\rm d}$$
$$T_{0} = 2451660.6432 \pm 0.0005$$
$$K_{2} = 144.6 \pm 0.3~ {\rm km s}^{-1}$$
 
\smallskip
\noindent
where $T_{0}$ is defined as the inferior conjunction of the optical star. 
The quoted uncertainties are 68 percent confidence and we have rescaled the 
errors so that the minimum reduced $\chi^{2}$ is 1.0.
Fig. 1 presents our radial velocity points folded on the best fitted solution.  
We have also extracted velocities using our green spectra 
but the orbital elements obtained are identical 
except for a small shift of +2 km s$^{-1}$ in the $\gamma$-velocity. 
Note that the 
velocity semiamplitude of the donor is extremely well constrained because we 
have concentrated our observations around the quadrature phases to minimize 
orbital smearing. Furthermore, the exquisite velocity accuracy of 
each data point ($\le$ 1 km s$^{-1}$) coupled with our long baseline leads 
to a determination of the orbital period with unprecedented accuracy.

\begin{figure}
\includegraphics[angle=0, scale=0.44]{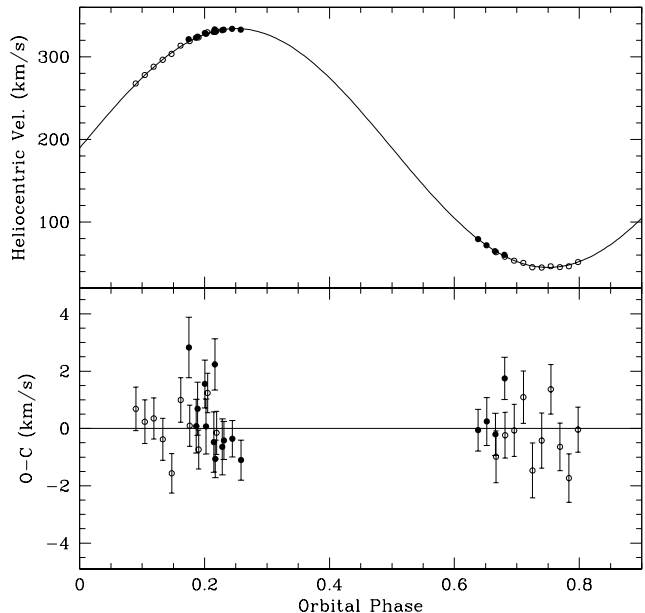}
\caption{Radial velocities of CenX-4 
folded on the orbital solution of Table 1 with best fitting sinusoid.
Individual velocity errors are always smaller than 1 km s$^{-1}$ and are 
shown in the bottom panel.  
Open circles represent the 2000 data and solid circles the 2004 data. 
\label{fig1}}
\end{figure}

We have also 
searched for evidence of period variations by comparing our T$_{0}$ with the
various spectroscopic zero phases reported in literature, which we list in 
Table 2. Given our long baseline we were able to estimate a zero phase for 
years 2000 and 2004 independently. Fig. 2 shows the (O-C) diagram for 
T$_0$=2449163.934 (Torres et al. 2002) and assuming a constant period at 
$P=0.6290522$  d. 
A quadratic fit provides a significantly better 
description of the (O-C) diagram than a linear fit (reduced $\chi^2$ of 
4.9 vs 9.7) and suggests that the orbital period of Cen X-4 decreases in a 
timescale of $P_{0}/\dot{P}=(1.22 \pm 0.31) \times 10^{6}$ yr. 
Note, 
however, that this claim relies on the very first epoch determination, 
obtained by Chevalier et al. (1989) through photometric lightcurves. 
If this is masked out a linear fit is marginally better and hence the 
evidence for a period variation vanishes. Therefore, this result must be 
treated with caution and further observations are needed to confirm or 
otherwise a possible period variation in Cen X-4. Furthermore, a period 
decrease would be unexpected from evolutionary grounds: conservation of 
angular momentum leads to an increase in binary separation and mass 
transfer is expected to be sustained by nuclear evolution of the companion 
(King, Kolb \& Burderi 1996).

\begin{figure}
\includegraphics[angle=0, scale=0.44]{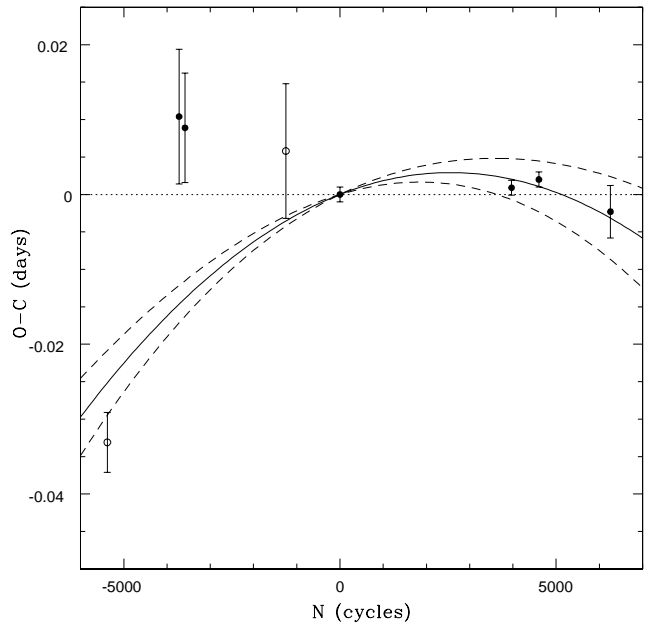}
\caption{The (O-C) with respect to the linear ephemeris with P=0.6290522 d 
and T$_{0}$=2449163.934. Open circles mark zero phases obtained photometrically 
and solid circles spectroscopically . The best quadratic fit is displayed, 
with the 1 $\sigma$ uncertainty level marked in dashed lines. 
\label{fig2}}
\end{figure}

\tabskip=0pt
\begin{table*}
\caption{Observed Zero Phases}
\label{tbl-2}
\begin{tabular}{cccl}
\hline
{Observed T$_{0}$} & {N} & 
{O-C} &  {Reference}\\
{(HJD-2440000)} & {(cycles)}& {(0.001d)}& \\
\hline
\hline
 5778.3420(40) & -5382 & -33.1 & Chevalier et al. (1989)\\
 6822.6121(90) & -3722 &  10.4 & Cowley et al. (1988) \\
 6909.4198(73) & -3584 &  8.9 & McClintock \& Remillard (1990) \\
 8376.3664(90) & -1252 &  5.8 & Shahbaz et al. (1993)\\
 9163.9340(10) &     0 &     0   & Torres et al. (2002) \\
11660.6431(10) &  3969 &  0.9 & VLT 2000 \\
12057.5761(10) &  4600 &  2.0 & D'Avanzo et al. (2005) \\
13099.9113(35) &  6257 & -2.3 & VLT 2004 \\
\hline
\hline
\end{tabular}
\end{table*}

We subsequently produced spectral averages in the rest frame of the companion 
star for the 2000 and 2004 databases, using our best ephemeris. Following 
Marsh, Robinson \& Wood (1994) we first computed the optimal $V_{\rm rot} \sin i$ by 
subtracting broadened versions of our synthetic template (in steps of 1 km 
s$^{-1}$) and minimizing the residual.  
We used a spherical rotational profile with linearized limb-darkening
$\epsilon=0.65$ and obtain $V_{\rm rot} \sin i$ = 46--47 km s$^{-1}$  for both 
databases, slightly higher but consistent within
1$\sigma$ with Torres et al. (2002). The error is likely 
to be dominated by the uncertainty in $\epsilon$ which is not known for the 
absorption lines (Collins \& Truax 1995, Shahbaz 2003). In an attempt to derive 
a more realistic error, we also computed the rotational broadening for the 
extreme cases where $\epsilon$=0--1 and find $V_{\rm rot} \sin i$ = 44--50 km 
s$^{-1}$ or 47$\pm$3 km s$^{-1}$. At this point we note that $V_{\rm rot} 
\sin i$ is known to be modulated with orbital phase because of tidal 
distortion of the donor star (e.g. see Casares et al. 1996). It shows a
double-humped oscillation with peak-to-peak amplitude of $\simeq$10\% and 
maxima at the quadratures, exactly 
when our data were taken. Therefore, our determination is likely to be
overestimated by $\simeq$5\% and hence we adopt $V_{\rm rot} \sin i = 44\pm 3$ km 
s$^{-1}$ for the sake of the system parameters determination. This value of 
the rotational broadening, combined with our refined $K_2$, implies a binary 
mass ratio $q=0.20 \pm 0.03 $ (Wade \& Horne 1988). We can now take our $q$ 
value into the ellipsoidal solutions of Shahbaz et al. (1993) and get a much 
better constraint on the inclination angle of $i=33-45^{\circ}$ (see their 
Fig. 3). This leads to a significant improvement in the binary masses, i.e. 
$M_{NS}= 1.14\pm 0.45$ \Msun~ and $M_{2}= 0.23\pm 0.10$ \Msun. The error 
budget is clearly dominated by the still large uncertainty in the inclination 
which needs to be refined through new high-quality ellipsoidal fits.

\section{Spectral Distribution}

Optimal subtraction of our broadened template from the 2000 and 2004 spectral 
averages yields relative contributions of the donor star to the 
total flux at $\lambda$6400 of 78 and 89 percent, respectively. This 
implies that the accretion disc is dimmer in 2004, by a factor 
$\simeq$ 2.3, with respect to 2000. In addition, the mean EW of the 
$H_{\alpha}$ 
emission drops from $\simeq 40$ \AA\ in 2000 to $\simeq 22$ \AA\ in 2004. This 
in turn implies that the emission line flux  also decreases by a factor 
$\simeq$ 2.1 between the two datasets. Significant variability in the accretion 
disc luminosity of Cen X-4 has been reported in previous works, both at short 
(i.e. minutes, Zurita, Casares \& Shahbaz 2003) and secular (i.e. 
days, Chevalier et al. 1989) timescales. 
Furthermore, Cen X-4 is one of the brightest SXTs 
in quiescence, with $L_{\rm X}\simeq 4 \times 10^{32}$ erg s$^{-1}$  and 
important variability (Campana et al. 2004 and references therein). 
It is likely that the optical changes that we observe are triggered by 
reprocessing of the variable X-ray flux into the accretion disc.

Figure 3 presents the average spectra of Cen X-4 in 2004, plotted in 
$F_{\lambda}$ units and dereddened by E(B-V)=0.1 (Blair et al. 1984). 
The three spectral ranges have been rebinned into 1000 pixels and smoothed 
with a gaussian bandpass of 2 pixels. We also show our synthetic template, 
conveniently rescaled to 89 percent of the total flux at $\lambda$6400. 
The bottom panel displays the residual of the subtraction which 
represents the spectral distribution of the accretion disc light. The 
continuum is well described by a 
standard $f_{\lambda} \propto \lambda^{-\alpha}$ law with $\alpha=1.4 \pm 0.1$,
in agreement with previous studies (Chevalier et al. 1988, Shahbaz et al 1993). 
Furthermore, extrapolation to 2000 \AA\ yields an excellent agreement with 
HST UV fluxes reported by McClintock \& Remillard (2000).  
The spectral distribution that we find is flatter than the standard 
$\alpha=2.3$ predicted for viscously heated optically thick accretion discs 
and suggests that irradiation of the outer disc regions may be important even in
quiescence.  
On the other hand, the emission line fluxes yield the following Balmer
decrement $H_{\alpha}/H_{\beta}/H_{\gamma}/H_{\delta}$=2.64/1/0.53/0.28, with 
$F(H_{\beta})=(2.14\pm0.06) \times 10^{-15}$ ergs cm$^{-2}$ s$^{-1}$ 
\AA$^{-1}$. This compares very well with the standard case B recombination 
for $T_{\rm eff} \simeq 10^4$ K suggesting that, at the large densities 
typical of accretion  discs ($N_{\rm e} \ge 10^{13}$ cm$^{-3}$), emission 
lines are powered by photoionization (Drake \& Ulrich 1980). The same 
result was also obtained for the quiescent black hole SXT V404 Cyg (Casares 
et al. 1993). This in turn supports a scenario where the optical variability 
observed in both the continuum and line fluxes is triggered by reprocessing of 
the variable X-ray flux  in the accretion disc. 

\begin{figure}
\resizebox{\hsize}{!}
{\includegraphics[angle=-90, scale=0.48]{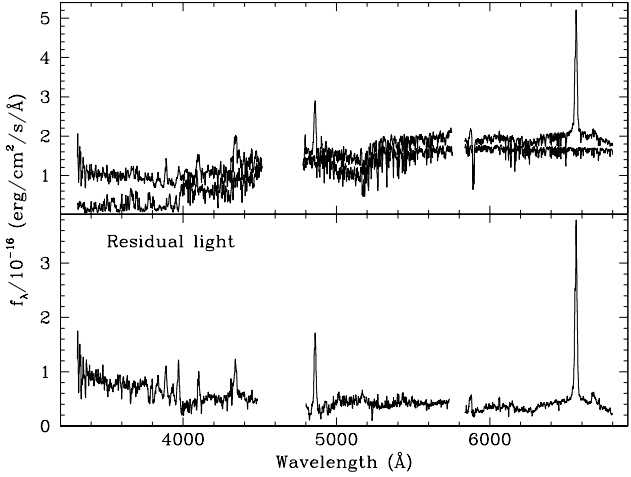}}
\caption{Top panel: Spectral distribution of Cen X-4, dereddened by 
$E_{B-V}=0.1$, and synthetic template, scaled to 89 percent of the total 
flux at $\lambda$6400. Bottom panel: spectral distribution of the accretion 
disc after subtracting the synthetic template from Cen X-4. 
\label{fig3}}
\end{figure}

\section{Li abundance and $^6$Li/$^7$Li isotopic ratio}

\subsection{Strategy of the analysis}

The D resonance doublets of $^6$Li and $^7$Li
are separated by about 0.16 \AA\ (the $^6$Li to the red), 
which implies that the $^6$Li D1 line is blended with the  $^7$Li 
D2 line. In practice these form a unique unresolved feature
even in warm slowly rotating stars, in which the feature 
is weak (see Smith et al. 1998 and Cayrel et al. 1999
for examples). The companion star of Cen X-4 is 
rotating rapidly and, to further complicate matters, the line
is strongly saturated.

To analyse the Li doublet we proceed in a manner similar
to that of Cayrel et al. (1999), who convincingly
argued that the Li feature should be fitted with
a synthetic spectrum using 5 parameters:
1) placement of the continuum, 2) and 3) abundances of
the two elements 4) macroscopic broadening of the lines
and 5) wavelength zero point adjustment.
The stars analysed by Cayrel at al. were all slow rotators,
thus the macroscopic broadening was dominated by
instrumental profile and macroscopic motions in the
atmosphere, in our case it is dominated by the large
rotation velocity. We thus used only $V_{\rm rot} \sin i$ as fitting
parameter, assuming a gaussian instrumental profile of 7 \kms, 
as deduced from the Th lines in the calibration arc.
Thus in our case the five parameters were:
1) placement of the continuum 2) $\log(N(^6 {\rm Li})/N({\rm Li_{tot}}))$
3) A(Li)\footnote{A(Li) = $\log (N({\rm Li})/N({\rm H}))+12$} 4)
$V_{\rm rot} \sin i$  5) wavelength zero point adjustment.
Of these 5 parameters 2 and 5 are correlated, quite obviously,
however also 2 and 3 are strongly correlated, as we shall show
below, due to the fact that the line is strongly saturated.

\subsection{Model atmosphere and synthetic spectra}

We used the same ATLAS LTE, plane parallel model atmosphere
as in Gonz\'alez Hern\'andez et al. (2005b), with $T_{\rm eff} = 4500$
K, $\log g = 3.9$ and $[\mathrm{M}/\mathrm{H}]=+0.25$. 
The lithium doublet is heavily saturated,
therefore we expect it to be seriously 
affected by microturbulence.
The determination of the microturbulent 
velocity requires the measurement of both
weak lines 
(on the linear part of the curve of growth)
and strong lines (on the flat part of the curve
of growth). For a rapidly rotating star,
like the companion to Cen X-4, the measurement
of the weak lines is impossible, therefore
one has no real handle on microturbulent
velocity. The need for a microturbulent
velocity arises from the limitations
of our one-dimensional model-atmospheres and the
need to take into account the velocity
fields which characterize a real stellar
atmosphere. If one were to use a full three
dimensional model atmosphere there should
be no need to introduce such a parameter.
It has long be recognized that microturbulence
is not a really independent parameter, but can
be calibrated as a function of effective temperature
and surface gravity. 
One such
calibration has been derived
for stars in the solar neighbourhood with similar
metallicity and stellar parameters (Allende 
Prieto et al. 2004), when applied to our
adopted atmospheric parameters 
it provides $\xi=0.91$ kms$^{-1}$.
It may be questionable whether
such a calibration is applicable 
to a stars which is rapidly rotating,
since it has been derived for ``normal'' slowly rotating
dwarfs of spectral types K to F. 
One could however argue that meridional 
flows induced by rotation should be rather
slow, for this rotational speed, and have
a character of {\em macrotrubulence}
rather than {\em microturbulence}, unless
the flow becomes unstable, producing small
scale disturbances. Therefore
we do not expect the microturbulence
to be too different from $\sim 1$  kms$^{-1}$.
Our adopted value is 
therefore $\xi=1$ kms$^{-1}$, however we shall
discuss the influence of microturbulence on our results. 
The synthetic spectra were computed with
the Linux version (Sbordone et al. 2005) of the SYNTHE suite (Kurucz 1993) 
We computed the emerging specific intensity for a mesh
of directions, then the emerging flux was obtained
by integrating the specific intensity and taking into
account the stellar rotation as described e.g. in Gray (1992)
and implemented by the code ROTATE of Kurucz (1993).
In this way limb-darkening is automatically taken into account.
We computed a small grid of synthetic spectra 
with 3 isotopic ratios, 3 Li abundances and 3 rotational 
velocities,  namely A(Li)=2.35
, 2.55 and 3.15, $\log(N(^6 {\rm Li})/N({\rm Li_{tot}}))$=-4, -1.125 and -0.301 
and $V_{\rm rot} \sin i$=40, 46 and 52 \kms. We used the same line list as in 
Gonz\'alez Hern\'andez et al. (2005b) plus the complete list of Kurucz (1993) 
and the TiO bands from Schwenke (1998).  
For Li we took into account the full HFS structure
listed by Kurucz (1995) as updated by Kurucz (2006). 

\begin{figure}
\centering
\includegraphics[angle=90, scale=0.32]{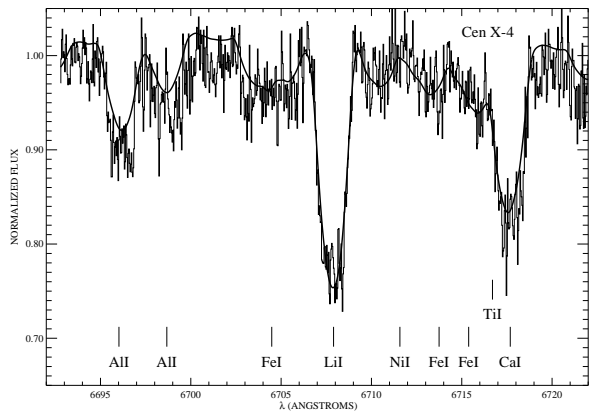}
\caption{Best fit to the average spectrum of year 2000, after
correcting every individual spectrum from the orbital solution presented 
in Sect. 3. A veiling factor of 0.22 (defined as the relative contribution 
of the disc to the total flux) was corrected from the observed spectrum.}              
\label{lifit}
\end{figure}

\begin{figure}
\centering
\includegraphics[angle=90, scale=0.32]{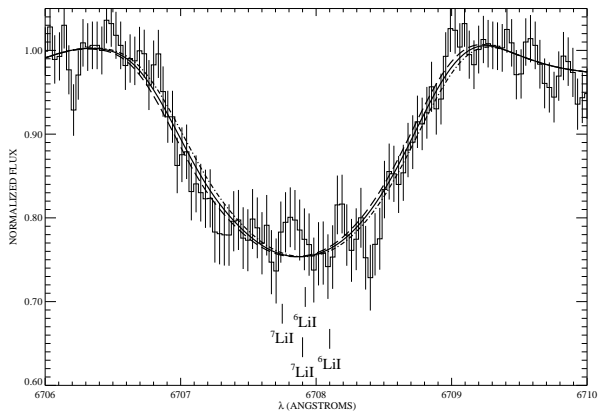}
\caption{Comparison of different fits to the lithium doublet, with
fixed  zero shift. Solid line: our best fit with all other 4 
parameters free which yields $\rm
\log(N(^6Li)/N(Li_{tot}))=-1.12$. Dashed line: fit with $\rm
\log(N(^6 Li)/N(Li_{tot}))=-3.5$ fixed. Dashed-dotted line: same with
$\rm \log(N(^6 Li)/N(Li_{tot}))=-0.47$ fixed. The spectrum of CenX-4
was corrected from a veiling factor of 0.22.} 
\label{varfit}
\end{figure}
 
\begin{table*}
\begin{minipage}[t]{\textwidth}
\caption{Best fits for different initial parameters}
\label{tbl-4}
\renewcommand{\footnoterule}{}  
\begin{tabular}{lccccccc}
\hline
\noalign{\smallskip}
Veiling & $\xi$(\kms) & Continuum & Shift(\kms) & $V_{\rm rot} \sin i$(\kms) & ${\rm A(Li)}$ & 
${\log(N(^6 {\rm Li})/N({\rm Li_{tot}}))}$ & ${N(^6 {\rm Li})/N(^7 {\rm Li})}$\\
\noalign{\smallskip}
\hline
\noalign{\smallskip}
\multicolumn{8}{c}{Single fits\footnote{Numbers in bold fonts
correspond to parameters that have been fixed}} \\
\noalign{\smallskip}
\hline
\hline
\noalign{\smallskip}
0.22 & 1 & 1.07 & 0.36  & 49.6 & 2.97 & -1.94 & 0.01    \\
0.22 & 1 & 1.07 & 0 & 49.6 & 2.87 & -1.12 & 0.08    \\
0.28 & 1 & 1.07 & 0 & 49.3 & 3.12 & -1.55 & 0.03    \\
0.22 & 2 & 1.07 & 0 & 50.1 & 2.78 & -0.97 & 0.12    \\
0.22 & 2 & 1.07 & 0 & 49.3 & 3.05 &  -4 &  $10^{-4}$ \\
\noalign{\smallskip}
\hline
\noalign{\smallskip}
\multicolumn{8}{c}{Monte Carlo Simulations} \\
\noalign{\smallskip}
\hline
\hline
\noalign{\smallskip}
0.22 & 1 & $1.07 \pm 0.01$ &  0 & $49 \pm 1$ & $2.87 \pm 0.20$ &
$-0.97 \pm 0.19$ & $0.12^{+0.08}_{-0.05}$\\
\noalign{\smallskip}
0.22 & 1 & $1.06 \pm 0.01$ & ${-0.42\pm 0.62}$ & $49 \pm 1$ & 
$2.83 \pm 0.12$ & $-1.05 \pm 0.22$ & $0.10^{+0.07}_{-0.04}$ \\
\noalign{\smallskip}
0.22 & 2 & $1.07 \pm 0.01$ & 0 & $50 \pm 1$ & $2.77 \pm 0.17$ &
$-0.91 \pm 0.22$ & $0.14^{+0.12}_{-0.06}$\\ 
\noalign{\smallskip}
0.22 & 2 & $1.07 \pm 0.01$ & ${-0.38\pm 0.50}$ & $50 \pm 1$ & 
$2.73 \pm 0.15$ & $-0.88 \pm 0.48$ & $0.15^{+0.51}_{-0.11}$\\ 
\noalign{\smallskip}
0.28 & 1 & $1.07 \pm 0.01$ & 0 & $49 \pm 1$ & 
$3.10 \pm 0.22$ & $-1.01 \pm 0.16$ & $0.11^{+0.05}_{-0.04}$\\ 
\noalign{\smallskip}
0.28 & 1 & $1.07 \pm 0.01$ & ${-0.77\pm 0.70}$ & $49 \pm 1$ & 
$3.01 \pm 0.14$ & $-1.02 \pm 0.22$ & $0.10^{+0.08}_{-0.04}$\\ 
\noalign{\smallskip}
\hline
\hline
\end{tabular}
\end{minipage}
\end{table*}

\begin{figure}
\centering
\resizebox{\hsize}{!}{\includegraphics[clip=true]{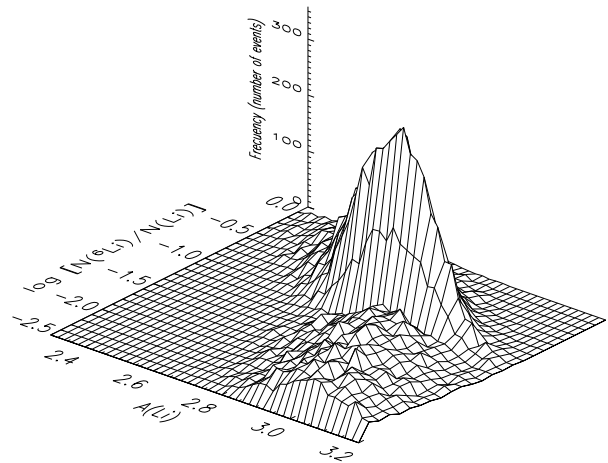}}
\caption{3-D view of the histogram of Monte Carlo events with
respect to Li isotopic ratio and total Li abundance for the model with
$\xi=1$ \kms, Shift$=0$ \kms (fixed) and veiling of 0.22.}
\label{MCper}
\end{figure}

\subsection{Line profile fitting}

We used as main data set the spectra observed in 
year 2000, since they have higher
signal-to-noise ratio  
with respect to those observed in year 2004 (73 vs 34).
Fits on the latter data provide, nevertheless, consistent results.
We unveiled our observed spectrum by 22 percent, as 
derived in Gonz\'alez Hern\'andez et al. (2005b) for the Li region.

We wrote  a $\chi^2$ fitting code, in which the minimum
is sought numerically using {\tt MINUIT} (James 1998)
and the synthetic spectra were interpolated in our grid.
One useful feature of {\tt MINUIT } is that it
allows to fix any of the parameters, if required.

As a first try we 
fitted the whole region 6692.6-6723.5 \AA\, where the Li line is 
located, and our best fit to the coadded spectrum of year 2000 
is shown in Fig. \ref{lifit}. 
This fitting region is rather wide
and we next attempted to fit only the 
region 6705.7-6709.4 \AA~ and  we find that the 
result does not depend on including other lines except for the Li feature.
In Table \ref{tbl-4} we show the results obtained
for different values of the veiling factor, the
microtrubulent velocity and leaving all 5 parameters
free, or fixing at 0 \kms ~ the velocity shift.

The main  motivation for searching for solution
with a fixed 0 \kms  ~ velocity shift
is that our coadded spectrum has been obtained 
by shifting at zero velocity all the individual
spectra, to better than 0.1 \kms, according
to our error estimate, and hence we expect to
find a zero velocity shift. 
To test the validity of a null
velocity shift, we have performed the spectral fitting to 
the range $\lambda\lambda$6300--6520 \AA\ and, indeed, obtain a zero 
point shift of 0 \kms.  
In the second place we want to test theories
which predict production of $^6$Li in 
low mass X-ray binaries, a positive velocity shift
will tend to conceal any possible presence of
$^6$Li. Therefore a fit with a fixed 0 \kms ~
velocity shift is useful and its result
should be considered to be the ``maximum'' fraction
of $^6$Li compatible with the data.
Finally also the results of Monte Carlo simulations
(see next section) support a 0 \kms ~ velocity shift
as the most likely. 

Inspection of Table \ref{tbl-4} reveals that
our derived Li abundance differs from that of Gonz\'alez Hern\'andez
et al. (2005b). This is mainly due to the inclusion of the $^6$Li
component but also, to some extent, to the readjustment of the continuum 
and rotational broadening, the different adopted microturbulence and 
spectrum synthesis code. In fact, when one fixes the Li isotopic ratio at -4 in
logarithmic scale (i.e. virtually no $^6$Li) the A(Li) obtained is almost 
identical to the value reported in Gonz\'alez Hern\'andez et
al. (2005b). 

All the isotopic ratios provided in Table \ref{tbl-4} 
imply very little $^6$Li
(recall that the meteoritic isotopic ratio is 
$\log(N(^6 {\rm Li})/N({\rm Li_{tot}}))=-1.123$).
The best fit, with 1 \kms ~ microturbulence,
0.22 veiling and fixed 0 \kms ~ velocity shift
is equivalent to $^6$Li/$^7$Li $\simeq $ 8\% and is shown in Fig. \ref{varfit} as a 
solid line. It is remarkable that in this case 
the $^6$Li content is consistent with the meteoritic abundance. 

In order to get a visual impression of the change the Li isotopic
ratio makes to the profiles we also show in Fig. \ref{varfit} 
two extreme fitting profiles:  one with a very low $^6$Li
content $\log(N(^6 {\rm Li})/N({\rm Li_{tot}}))=-3.5$ is shown as a dashed line,
whereas another fit with very large $^6$Li content 
$\log(N(^6{\rm Li})/N({\rm Li_{tot}}))=-0.47$ is plotted as
dash-dotted line. These correspond to $^6$Li/$^7$Li = 0.03\% and 50\%
respectively. 

From Table \ref{tbl-4} we may also deduce the impact on the derived Li
abundance and isotopic ratio of the veiling factor and microturbulent
velocity. We used the estimate of the veiling factor uncertainty by 
Gonz\'alez Hern\'andez et al. (2005b).
For the reader's convenience the resulting uncertainties
in A(Li), logarithmic $^6$Li fraction and $^6$Li/$^7$Li ratio are
provided in Table \ref{tbl-3}, which has been estimated for the case
where the shift is fixed at 0 \kms.  
Since Gonz\'alez Hern\'andez et al.
(2005b) assumed a microturbulent velocity of 2 \kms, 
we have also quantified the effect of the
microturbulence on the metallicity. We found that the metallicity
decreases 0.1 dex for an increase of 0.25 \kms. This result is 
not surprising, since only strong Fe~I lines could be analysed due to the large
rotational velocity of the secondary star.

\begin{table}
\begin{minipage}[t]{\columnwidth}
\caption{Impact of microturbulence
and veiling factor on derived Li abundance and isotopic ratio}
\label{tbl-3}
\begin{tabular}{lccc}
\hline
\noalign{\smallskip}
Parameter & $\Delta_{\rm A(Li)}$ & 
$\Delta_{\log(N(^6 {\rm Li})/N({\rm Li_{tot}}))}$  &  
${N(^6 {\rm Li})/N(^7 {\rm Li})}$\\
\noalign{\smallskip}
\hline
\noalign{\smallskip}
\multicolumn{4}{c}{Single fits} \\
\noalign{\smallskip}
\hline
\hline
\noalign{\smallskip}
$\Delta_\xi= +1$ \kms & -0.09 & +0.15 & +0.04\\
$\Delta_{\rm veiling}= +0.06$ & +0.25 & -0.43 & -0.05 \\
\noalign{\smallskip}
\hline
\noalign{\smallskip}
\multicolumn{4}{c}{Monte Carlo Simulations} \\
\noalign{\smallskip}
\hline
\hline
\noalign{\smallskip}
$\Delta_\xi= +1$ \kms & -0.10 & +0.06 & +0.02 \\
$\Delta_{\rm veiling}= +0.06$ & +0.23 & -0.04 & -0.01 \\
\noalign{\smallskip}
\hline
\hline
\end{tabular} 
\end{minipage}
\end{table}

\begin{figure}
\centering
\resizebox{\hsize}{!}
{\includegraphics[clip=true]{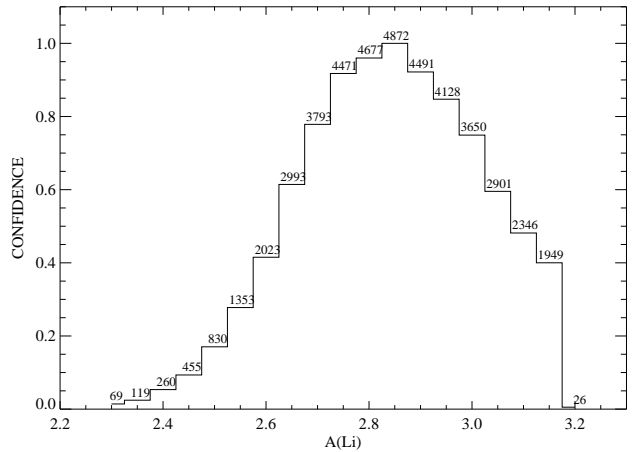}}
\caption{Histogram of Monte Carlo events with respect to total Li
abundance for the model with $\xi=1$ \kms, Shift$=0$ \kms and veiling
of 0.22.}
\label{mcali}
\end{figure}

\begin{figure}
\centering
\resizebox{\hsize}{!}
{\includegraphics[clip=true]{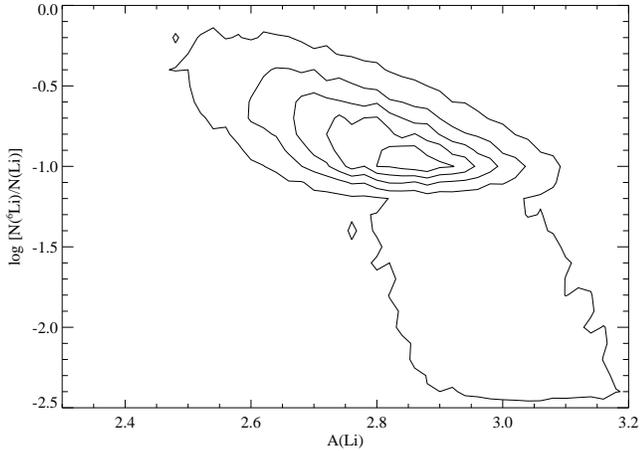}}
\caption{Contour plot of the histogram of Monte Carlo
events with respect to Li isotopic ratio and total Li abundance for
the model with $\xi=1$ \kms, Shift$=0$ \kms and veiling 
of 0.22.}
\label{contour}
\end{figure}

\subsection{Confidence intervals}

Although we have used $\chi^2$ fitting, it is clear
that all the theorems which allow to derive errors
and confidence limits
from $\chi^2$ do not apply to the case of
spectra, since adjacent pixels are 
always correlated, as pointed out by Cayrel et al. (1999)
and Caffau et al. (2005). 
However, confidence limits can be reliably estimated using
Monte Carlo techniques. To this end we
performed  Monte Carlo simulations by injecting 
Poisson noise, according to the standard deviations 
used in the fitting, into the best fitting synthetic
spectrum and performing the fit with the same code
used for the observed spectrum, for over 50000 
samples. 

The results of four such simulations are
given in the last six lines of Table \ref{tbl-4}. 
We would like to stress that the
Monte Carlo simulations yield no significant dependence of the
derived isotopic ratios on the values adopted for the microturbulent
velocity and veiling factor, and are all consistent within 1-$\sigma$.
The two dimensional
histogram with respect to isotopic ratio and A(Li) 
of the simulation with $\xi=1$ \kms ~ and
a fixed 0 \kms ~ velocity shift, is shown in Fig.
\ref{MCper}. The strong correlation of the two parameters is 
clear and results from the fact that the line is strongly 
saturated. The higher the $^6$Li fraction and 
the more the line is de-saturated, thus allowing to reach the 
observed equivalent width with a lower A(Li).  
In Fig. \ref{contour} a countour plot of the same histogram
is shown. From this distribution one may estimate the area which
encloses roughly 68\% of the Monte Carlo samples, 
which may be taken as akin to $1\sigma$ for normal distributions.
We find that the highest Li isotopic ratio 
compatible with the observations at $1\sigma$   
is $\log(N(^6 {\rm Li})/N({\rm Li_{tot}}))=-0.78$ which corresponds
to $N(^6{\rm Li})/N(^7{\rm Li})$ = 0.20. In fact 68\% of 
the events are bounded by the following limits $N(^6 {\rm
Li})/N({\rm Li_{tot}}) = 0.07-0.16$ and A(Li)=2.67--3.07.  
Therefore, we can claim a $1\sigma$ detection of the Li isotopic ratio 
in Cen X-4 with $N(^6{\rm Li})/N(^7{\rm Li})=0.12^{+0.08}_{-0.05}$ and 
$\rm A(Li)= 2.87\pm0.20$. The most likely value for the Li abundance
and its uncertainty was infered from the histogram extracted from the
Monte Carlo simulations which is shown in Fig. \ref{mcali}. 
Note that, although 
the most probable value of A(Li) of the distribution
coincides with the value found by the best single fit, the
corresponding isotopic ratio
 is slightly different.

It is interesting to note that in both the Monte Carlo simulations 
done leaving all five parameters free, the mean value of the
velocity shift is consisten with 0 \kms, within 1 standard deviation.
In our view this strongly argues in favour
of fixing at 0 \kms ~ the velocity shift.

\section{Adopted A(Li) and isotopic ratio}

All the results of our fits and Monte Carlo simulations 
are listed in Table \ref{tbl-4}, whereas Table~\ref{tbl-3} 
summarises the impact of the veiling and microturbulent velocity.

We have already discussed the issue of veiling
factor and microturbulent velocity and from our point
of view the best choice is 0.22 veiling factor
and 1 \kms microturbulent velocity.
Concerning the velocity shift we believe that
the results of the Monte Carlo simulation
and the process of optimal coaddition of the spectra
adopted, strongly suggest that this parameter
should be fixed at 0 \kms. 
We are left to chose between two possible couples of values
for A(Li) and 
$\log(N(^6 {\rm Li})/N({\rm Li_{tot}}))$: the best fit with a 
fixed 0 \kms ~ velocity shift or the result
of the corresponding Monte Carlo simulation.
It is reassuring that the value of A(Li) is the same
in both cases. The difference in isotopic ratio
is small and boils down to $^6$Li/$^7$Li = 8\%
for the direct fit or $^6$Li/$^7$Li=12\% from
the Monte Carlo simulation.
Our discussion would not change if we adopted one or the
other. However we decided to adopt the result
of the Monte Carlo simulation. We believe that in a situation
like this, with a minimum with a complex topology,
and strongly correlated parameters, the Monte Carlo
simulation does a better job in estimating the most
likely parameters.
  
We also note that $<4$\% of the events have $N(^6 {\rm Li})/N({\rm
Li_{tot}})>0.4$ which defines a 3-$\sigma$ upper limit to the $^6$Li
abundance of $N(^6{\rm Li})/N(^7{\rm Li})<0.66$.   

\section{Discussion}

The Li abundance in the companion of Cen X-4 is known 
to be  high  (Mart\'\i{}n et al 1994, Gonz\'alez Hern\'andez et al. 
2005b), and is confirmed by the present result of A(Li)= 2.87. 
The relevant aspect is that this value is higher of what expected 
for a cool K star, in which the convective zone is deep enough to destroy 
the surface Li. 
This characteristic is shared by other companions in X-ray binaries. 
Li detections have been reported in the companion stars to several 
SXTs, namely V404 Cyg (Mart\'\i{}n et al. 1992), A0620-00 (Marsh et al. 1994),  
GS 2000+25 (Filippenko, Matheson \& Barth 1995)   
and Nova Mus (Mart\'\i{}n et al. 1996). 

\tabskip=0pt
\begin{table*}
\begin{minipage}[t]{15cm}
\caption{Li detection in SXTs}
\label{tbl-5}
\renewcommand{\footnoterule}{}  
\begin{tabular}{lccccccc}
\hline
{System} &  {$V_{\rm rot} \sin i$} & {i} &  {$V_{\rm rot}$} & {Spectral} & 
{EW Li} & {A (Li)} & {Reference\footnote{(1) Orosz 2001. (2) Gelino et al. 2006. (3) Harlaftis et al. 1999. 
(4) Gelino \& Harrison 2003. (5) Harlaftis et al. 1997. (6) Remillard et al. 1996. (7) Harlaftis et al. 1996. 
(8) Callanan et al. 1996. (9) Casares et al. 1997. (10) Gelino, Harrison \& McNamara 2001 (11) Marsh, Robinson 
\& Wood 1994. (12) Gelino, Harrison \& Orosz 2001 (13) Casares \& Charles 1994 (14) Shahbaz et al. 1994 
(15) This paper. (16) Shahbaz, Naylor \& Charles 1993.}} \\
 & {(\kms)} & {(deg.)} & {(\kms)} & {Type} & {(\AA)} & & \\ 
\hline
\hline
J1118+480 & 114 $\pm$4       & 68$\pm$2  & 123$\pm$5   &   K5-7VI & $<$0.16          & $<1.8$          & (1,2) \\ 
J0422+32  & 90$^{+22}_{-27}$ & 45$\pm$2  & 127$\pm$36  &   M2-4V  & $<$0.21          & $<1.62$	       & (3,4) \\
N. Oph 77 & 50$^{+17}_{-23}$ & 70$\pm$10 &  53$\pm$23  &   K5-7IV & $<$0.37          & $<2.96$         & (5,6) \\
GS2000+25 &  86 $\pm$8       & 65$\pm$9  &  95$\pm$11  &   K3-6V  &  0.150$\pm$0.090 & 2.20 $\pm$ 0.50 & (7,8) \\
GU Mus    & 106 $\pm$13      & 54$\pm$2  & 131$\pm$16  &   K3-K4V &  0.420$\pm$0.060 & 3.00 $\pm$ 0.50 & (9,10) \\
V616 Mon  &  93 $\pm$4       & 41$\pm$3  & 142$\pm$11  &   K2V	  &  0.245$\pm$0.030 & 2.31 $\pm$ 0.21 & (11,12) \\
V404 Cyg  &  41 $\pm$1       & 55$\pm$4  &  50$\pm$3   &   K0IV   &  0.290$\pm$0.030 & 2.70 $\pm$ 0.40 & (13,14) \\
Cen X-4   &  44 $\pm$3       & 40$\pm$7  &  68$\pm$11  &   K4IV   & 
0.440$\pm$0.025 & 2.87\footnote{Note that in this case the Li abundance has been
derived taking into account the presence of $^6$Li, at variance with what done for the other stars.
A fit to the data assuming no $^6$Li would provide an abundance higher by 0.12 dex.} $\pm$ 0.20 & (15,16) \\
\hline
\hline
\end{tabular}
\end{minipage}
\end{table*}
\normalsize

The surprising high abundance of Li  observed in the cool companions of 
compact objects has been taken as an evidence of local Li production
connected to the presence of relativistic particles around the compact objects.
These are binary systems in which a compact object, a neutron star or a black 
hole, is accreting matter from the companion and Li production can occur either 
in the accretion flow or on the surface of the companion, but through processes
associated with the accretion (see Guessoum  \&  Kazanas  1999 and references 
therein). Proposed physical mechanisms are $\alpha-\alpha$ fusion in the 
accretion flow (Yi \& Narayan 1997) or spallation of CNO elements 
by neutrons, on the star surface (Guessoum \& Kazanas 1999).
This leads to the prediction that other light nuclei  such as 
$\rm ^6 Li$, Be and B  must be synthesized  besides 
$\rm ^7 Li$, the amount depends on the details of the accretion process.

However,  Li in these systems is never found in excess of the presently cosmic, 
i.e. meteoritic value, A(Li)=3.3 which would be an unambiguous signature of 
fresh Li production.  Thus it has also been suggested that instead 
of Li production we are dealing with the supression of  the destruction 
mechanism, which is  normally very efficient for late type stars. Theoretically 
rotation has been shown to counteract  Li depletion and some evidence of 
reduced Li depletion has been found in tidally locked binaries in open clusters.
Mart{\'\i}n \& Claret (1996) showed that high enough rotational 
velocities may inhibit Li depletion in the atmospheres of such 
late-type stars. On the contrary, Pinsonneault et al. (1990) have 
found that the effects of high rotational velocities on the Li 
depletion processes are small and these effects are not resposible for 
the Li abundance dispersion in Pleiades K-dwarfs with similar 
effective temperatures and masses, discarding a clear correlation 
between non-projected rotational velocities (obtained from 
photometric periods) and Li abundances (King et al. 2000). More recently,   
Maccarone, Jonker \& Sills (2005) have argued that inhibited 
Li destruction, due to tidally locked rotation of the companion stars, 
could explain the large abundances observed in SXTs. X-ray binaries 
spend several Gyr of their lifetimes as tidally locked binaries 
and it is only when they come into the Roche-lobe contact that 
some Li depletion, through mass transfer, starts. Cataclysmic variables, which 
are systems related to the SXT, do not show the presence of Li. This fact has 
been taken as evidence of Li production in the SXT (Mart\'\i{}n et 
al. 1995). However, Maccarone et al pointed out that cataclysmic variables 
have longer lifetimes compared with the systems ending with a black hole or a 
neutron star  and that the  companions in cataclysmic variables are not 
tidally locked for most of their  evolution prior to the Roche lobe contact 
phase. These differences could explain the non detection of Li in the  
companions of cataclysmic variables. 

In Table 5 we list the EWs (corrected for veiling of the accretion
disc) and Li abundances of several SXTs, together with the stellar 
rotation $V_{\rm rot}$, binary inclination (based on 
ellipsoidal model fits) and spectral type of their 
companion stars, as reported in the papers listed in column 8. 
A(Li) values are taken from the following  
works: Mart\'\i{}n et al. (1996) for V404 Cyg, GU Mus and GS2000+25;  
Gonz\'alez Hern\'andez et al. (2004) for A0620-00 and  Gonz\'alez Hern\'andez 
et al. (2006) for XTE J1118+480. Upper limits to A(Li) have also been computed 
for GRO J0422+32 and N. Oph 77 through the code MOOG (Sneden 1973)
using the reported EWs and $T_{\rm eff}$ of companion stars. 
Note that Li abundances in this table do not consider $^6$Li, except for the
case of Cen X-4.  
$V_{\rm rot}$ is calculated from the measured rotational broadening 
$V_{\rm rot} \sin i$ and the inclination. Table 5 shows that there is no obvious 
correlation between A(Li) and $V_{\rm rot}$ but do seems to be a
connection with the spectral type of  the star: Li is only detected in
companions with spectral type earlier than K5, suggesting that the
depth of the convection layer seems to be an important 
factor. This can be taken as circumstantial support to preservation
models.  

A measurement to the $\rm ^6 Li/ ^7 Li $ 
ratio can, in principle, discriminate between production or 
preservation scenarios.
This is because  $\alpha$ + $\alpha$ fusion reactions occurring in the inner
region of an accretion disc containing helium  would produce  significant 
amount of $^6$Li beside $^7$Li and $^7$Be which decays into $^7$Li with a 
lifetime of 76 days. 
Unfortunately, there are no computations of $\alpha$ + $\alpha$ reactions 
in accretion discs with realistic (solar) abundances.   
Guessoum et al. (1997) examined the nuclear reactions that 
occur in the inner regions of accretion discs composed of pure helium. They 
computed the gamma-ray lines  resulting from $^7$Li and $^7$Be deexcitation 
lines which produce a feature at about 0.450 MeV and, incidentally,  
might have produced the 0.5 Mev emission feature observed in Nova Muscae. 
They conclude that the Li isotopic ratio is expected to be in the range 
$0.1 < N(^6{\rm Li})/N(^7{\rm Li}) < 10$ for ion temperatures of the plasma varying 
from KT$_{i}$=3MeV upto 50 MeV (Guessoum et al. 1997).
Our analysis shows that $N(^6{\rm Li})/N(^7{\rm Li})=0.12^{+0.08}_{-0.05}$ and is consistent with the
computations for ion temperatures $\sim 4.3$ MeV. Therefore, 
our value cannot rule out Li production models, 
however we stress that the low level of
$^6$Li is remarkable. This, together with 
the possible correlation between A(Li) and spectral types and the 
non detection of a single X-ray binary with supermeteoritic A(Li), seems to
support Li preservation scenarios. 
We also note that a new evolutionary picture has recently been proposed
by Ivanova (2006) where LMXBs are powered by mass transfer from pre-main
sequence donors. This would help to explain the roughly primordial abundance of
Li in Cen X-4 through youth and, hence, our value of the isotopic ratio. 
Clearly, in  order to confirm our result, 
new computations of $\alpha$ + $\alpha$ fusion and spallation reactions in 
accretion discs with solar composition are required.

\begin{acknowledgements}

Use of  the MOLLY and DOPPLER
software  developed by T. R.  Marsh  is gratefully acknowledged. 
We are grateful to H.-G. Ludwig for 
interesting discussions on the role of microturbulence
and its possible connection to rotation.
PB and PM acknowledge support from MIUR - PRIN grant 2004025729
(P.I. M. Busso). PB and JIGH acknowledge support from EU
contract MEXT-CT-2004-014265 (CIFIST). JC akcnowledges support by Spanish MCYT 
grant AYA2002-0036.
\end{acknowledgements}



\begin{thebibliography}{}

\bibitem[]{}Allende Prieto, C., Barklem, P. S., Lambert, D. L., Cunha,
K. 2004, A\&A, 420, 183
\bibitem[]{}Blair W.P., Raymond J.C., Dupree A.K., Wu C.C., Holm A.V., 
Swank J.H. 1984, ApJ, 278, 270
\bibitem[]{}Caffau E., Bonifacio P., Faraggiana R., Fran\c cois P.,
Gratton R.G., Barbieri M., 2005 \aap, 441, 533
\bibitem[]{}Callanan P.J., Garcia M.R., Filippenko A.V., McLean I., Teplitz H. 1996, ApJ, 470, L57
\bibitem[]{}Campana, S., Israel G.L., Stella L., Gastaldello F., Mereghetti S. 
2004, ApJ, 601, 474
\bibitem[]{}Casares J. 2001 in {\it Binary stars: Selected Topics
on Observations and Physical processes}, eds. L\'azaro, F.C. \&
Ar\'evalo, M.J., Lecture Notes in Physics, Vol. 563, p.277
\bibitem[]{}Casares J. 2005 in {\it The Many Scales of the Universe - 
JENAM 2004 Astrophysics Reviews}, eds. del Toro Iniesta J.C. et al., 
Kluwer Academic Publishers, in press (astro-ph/0503071)
\bibitem[]{}Casares J., Charles P. A. 1994, MNRAS, 271, L5
\bibitem[]{}Casares J., Charles P. A., Naylor T., Pavlenko E. P. 1993, 
MNRAS, 265, 834
\bibitem[]{}Casares J., Mart\'\i{}n E.L., Charles P. A., Molaro P., Rebolo R. 
1996, NewA, 1, 299
\bibitem[]{}Casares J., Mouchet M., Mart\'\i{}nez-Pais I.G., Harlaftis E.T. 
1996, MNRAS, 282, 182
\bibitem[Cayrel et al.(1999)]{1999A&A...343..923C} Cayrel, R., Spite, M., 
Spite, F., Vangioni-Flam, E., Cass{\' e}, M., \& Audouze, J.\ 1999, \aap, 
343, 923 
\bibitem[]{}Charles P.A., Coe M.J. 2006 in {\it Compact Stellar X-ray 
Sources}, eds. Lewin W.H.G. \& van der Klis M., Cambridge Astrophysics Series No
39, CUP, p.215
\bibitem[]{}Chevalier C., Ilovaisky S.A., van Paradijs J.,
Pedersen H., van der Klis M. 1989, A\&A, 210, 114
\bibitem[]{}Collins G.W.II, Truax R.J. 1995, ApJ, 439, 860
\bibitem[]{}Conner J.P., Evans W.D., Belian R.D. 1969, ApJ, 157, L157
\bibitem[]{}Cowley A.P., Hutchings J.B., Schmidtke P.C., Hartwick F.D.A.,
Crampton D., Thompson I.B. 1988, AJ, 95, 1231
\bibitem[]{}D'Avanzo P., Campana S., Casares J., Israel G.L., Covino S., Charles
P.A., Stella L. 2005, A\&A, 444, 905
\bibitem[]{}Drake S.A., Ulrich R.K. 1980, ApJS, 42, 351 
\bibitem[]{}Eggleton P.P., Bailyn C.D., Tout, C.A. 1989, ApJ, 345, 489
\bibitem[]{}Filippenko A.V., Matheson Th., Barth A.J. 1995, ApJ, 455, L139
\bibitem[]{}Garcia M.R., Bailyn C.D., Grindlay J.E., Molnar
L.A. 1989, ApJ, 341, L75
\bibitem[]{}Gelino D.M., Harrison T.E., McNamara B.J. 2001, AJ, 122, 971
\bibitem[]{}Gelino D.M., Harrison T.E., Orosz J.A. 2001, AJ, 122, 2668
\bibitem[]{}Gelino D.M., Harrison T.E. 2003, ApJ, 599, 1254
\bibitem[]{}Gelino D.M. et al. 2006, ApJ, 642, 438
\bibitem[]{}Gonz\'alez Hern\'andez J.I., Rebolo R., Israelian G., 
Casares J., Maeder A., Meynet G. 2004, ApJ, 609, 988 
\bibitem[]{}Gonz\'alez Hern\'andez J.I., Rebolo R., Pe\~narrubia J., 
Casares J., Israelian G. 2005a, A\&A, 435, 1185
\bibitem[]{}Gonz\'alez Hern\'andez J.I., Rebolo R., Israelian G., 
Casares J., Maeda K., Bonifacio P., Molaro P. 2005b, ApJ, 630, 495 
\bibitem[]{}Gonz\'alez Hern\'andez J.I., Rebolo R., Israelian G., 
Harlaftis E.T., Filippenko A.V., Chornock R. 2006, ApJ, 644, L49 
\bibitem[]{}Gray D.F. 1992, "The observation and Analysis of Stellar
Photospheres", CUP 20
\bibitem[]{}Greenstein J.L., Saha A. 1986, ApJ, 304, 721
\bibitem[]{}Guessoum N., Kazanas D., Kozlovsky, B. 1997, Proceedings of the 
25th International Cosmic Ray Conference, Durban, South Africa, July 30 - 
August 6, 1997, edited by M.S. Potgieter et al. (Wesprint, Potchefstroom, 
Space Research Unit, 1997), p. 145
\bibitem[Guessoum \& Kazanas(1999)]{1999ApJ...512..332G} Guessoum N.,  
Kazanas D.\ 1999, \apj, 512, 332 
\bibitem[]{}Harlaftis E. T., Horne K., Filippenko A. V. 1996, PASP, 108, 762
\bibitem[]{}Harlaftis E. T., Steeghs D., Horne K., Filippenko A. V. 1997, AJ, 114, 1170
\bibitem[]{}Harlaftis E. T., Collier S., Horne K., Filippenko A. V. 1999, A\&A, 341, 491
\bibitem[]{}Israelian G., Rebolo R., Basri G., Casares J., Mart\'\i{}n E.L. 
1999, Nature, 401, 142
\bibitem[]{}Ivanova N. 2006, ApJ, 653, L1371  
\bibitem[{James (1998)}]{minuit}
James, F., 1998
MINUIT, Reference Manual, Version 94.1, CERN, Geneva, Switzerland
\bibitem[]{}Kaluzienski L.J., Holt S.S. Swank J.H,. 1980, ApJ, 241, 779
\bibitem[]{}King A.R., Kolb U., Burderi L. 1996, ApJ, 464, L127
\bibitem[]{}King, J. R., Krishnamurthi, A., \& Pinsonneault, M. H. 2000, ApJ, 
119, 859
\bibitem[{Kurucz(1993)}]{kurucz93}Kurucz R. L., 1993, CDROM 13, 18
\bibitem[Kurucz(1995)]{1995ApJ...452..102K} Kurucz, R.~L.\ 1995, \apj, 452, 
102 
\bibitem[{Kurucz(2006)}]{}Kurucz R. L., 2006, http://kurucz.harvard.edu/atoms/0300
\bibitem[]{}Maccarone T.J., Jonker P.G., Sils A.I. 2005, A\&A, 436, 671
\bibitem[]{}Marsh T. R., Robinson E. L., Wood J. H. 1994, MNRAS 266, 137
\bibitem[]{}Mart\'\i{}n E.L., Rebolo R., Casares J., Charles P.A. 1992, 
Nature, 358, 129
\bibitem[]{}Mart\'\i{}n E.L., Rebolo R., Casares J., Charles P.A. 1994, 
ApJ, 435, 791
\bibitem[]{}Mart\'\i{}n E.L., Casares J., Charles P.A., Rebolo R.  1995, 
A\&A, 303, 785
\bibitem[]{}Mart\'\i{}n E.L., Casares J., Molaro P., Rebolo R., 
Charles P.A. 1996, NewA, 1, 197
\bibitem[]{}Mart{\'\i}n, E. L., \& Claret, A. 1996, A\&A, 306, 408 
\bibitem[]{}McClintock J.E.,  Remillard R.A. 1990, ApJ, 350, 386
\bibitem[]{}McClintock J.E.,  Remillard R.A. 2000, ApJ, 531, 956
\bibitem[]{}Orosz J.A. 2001, Astron. Telegram, 67, 1
\bibitem[]{}Pinsonneault, M. H., Kawaler, S. D., \& Demarque, P. 1990, ApJS, 74, 501 
\bibitem[]{}Remillard R.A., Orosz J.A., McClintock J.E., Bailyn C.D. 1996, ApJ, 459, 226
\bibitem[{Sbordone et al. (2005) }]{sbordone}
Sbordone, L., Bonifacio, P., Castelli, F., \& Kurucz, R. L., 2004, 
MSAIS, 5, 93 
\bibitem[]{}Schwenke D.M. 1998, Chemistry and Physics of Molecules and Grains in
Space. Faraday Discussions No. 109. The Faraday division of the Royal Society of
Chemistry, London, p. 321  
\bibitem[]{}Shahbaz T., Naylor T., Charles P.A. 1993, MNRAS, 265, 655
\bibitem[]{}Shahbaz T., Ringwald F.A., Bunn J.C., Naylor T., Charles P.A., Casares J. 
1994, MNRAS, 271, L10
\bibitem[]{}Shahbaz T. 2003, MNRAS, 339, 1031
\bibitem[Smith et al.(1999)]{1999ApJ...516L..73S} Smith, V.~V., Shetrone, 
M.~D., \& Keane, M.~J.\ 1999, \apjl, 516, L73 
\bibitem[]{}Sneden C. 1973, Ph.D. thesis, Univ. Texas, Austin
\bibitem[]{}Torres M.P., Casares J., Mart\'\i{}nez-Pais I.G., 
Charles P.A. 2002, MNRAS, 334, 233
\bibitem[]{}van Paradijs J., McClintock J.E. 1995, in {\it X-ray
Binaries}, eds. Lewin, W.H.G., van Paradijs, J. \& van den Heuvel, E.P.J.
CUP, p.58
\bibitem[]{}Wade R.A., Horne K. 1988, ApJ, 324, 411
\bibitem[]{}Yi I., Narayan R. 1997, ApJ, 486, 363
\bibitem[]{}Zurita C., Casares J., Shahbaz T. 2003, ApJ, 582, 369

\end{thebibliography}
\end{document}